\documentclass[conference]{IEEEtran}
\IEEEoverridecommandlockouts

\usepackage{cite}
\usepackage{amsmath,amssymb,amsfonts}
\usepackage{graphicx}
\usepackage{textcomp}
\usepackage{xcolor}

\usepackage{algorithm}
\usepackage{algpseudocode}

\usepackage{float}
\usepackage[top=0.75in, bottom=1.05in, left=0.625in, right=0.625in]{geometry}

\def\BibTeX{{\rm B\kern-.05em{\sc i\kern-.025em b}\kern-.08em
    T\kern-.1667em\lower.7ex\hbox{E}\kern-.125emX}}
\begin{document}

\title{An Automated, Scalable Machine Learning Model Inversion Assessment Pipeline}

\makeatletter
\newcommand{\linebreakand}{%
  \end{@IEEEauthorhalign}
  \hfill\mbox{}\par
  \mbox{}\hfill\begin{@IEEEauthorhalign}
}
\makeatother
 
\author{
    \IEEEauthorblockN{Tyler Shumaker, Jessica Carpenter,\\ David Saranchak}
    \IEEEauthorblockA{\textit{Concurrent Technologies Corporation}\\
    Johnstown, PA, 15904 USA \\
    \{shumaket,carpentj,saranchd\}@ctc.com}
\and
    \IEEEauthorblockN{Nathaniel D. Bastian}
    \IEEEauthorblockA{\textit{United States Military Academy}\\
    West Point, NY, 10996 USA \\
    nathaniel.bastian@westpoint.edu}
}

\maketitle

\begin{abstract}
Machine learning (ML) models have the potential to transform military battlefields, presenting a large external pressure to rapidly incorporate them into operational settings. However, it is well-established that these ML models are vulnerable to a number of adversarial attacks throughout the model deployment pipeline that threaten to negate battlefield advantage. One broad category is privacy attacks (such as model inversion) where an adversary can reverse engineer information from the model, such as the sensitive data used in its training. The ability to quantify the risk of model inversion attacks (MIAs) is not well studied, and there is a lack of automated developmental test and evaluation (DT\&E) tools and metrics to quantify the effectiveness of privacy loss of the MIA. The current DT\&E process is difficult because ML model inversions can be hard for a human to interpret, subjective when they are interpretable, and difficult to quantify in terms of inversion quality. Additionally, scaling the DT\&E process is challenging due to many ML model architectures and data modalities that need to be assessed. In this work, we present a novel DT\&E tool that quantifies the risk of data privacy loss from MIAs and introduces four adversarial risk dimensions to quantify privacy loss. Our DT\&E pipeline combines inversion with vision language models (VLMs) to improve effectiveness while enabling scalable analysis. We demonstrate effectiveness using multiple MIA techniques and VLMs configured for zero-shot classification and image captioning. We benchmark the pipeline using several state-of-the-art MIAs in the computer vision domain with an image classification task that is typical in military applications. In general, our innovative pipeline extends the current model inversion DT\&E capabilities by improving the effectiveness and scalability of the privacy loss analysis in an automated fashion. With it, a clearer understanding of the risk to military ML-enabled capabilities can be obtained.
\end{abstract}


\section{INTRODUCTION}

Military battlefields are highly complex and dynamic environments, where small imbalances in technological capabilities can provide significant advantages. Operational technological systems that employ artificial intelligence (AI) are expected to provide new advantages in operational capabilities and decision-making abilities as their maturity increases within the next decade. These military AI systems depend upon machine learning (ML) techniques that require proprietary training data, often acquired at significant cost. However, there are significant risks to these deployed systems. A deployed ML model could be captured, exposing its architecture, weights, and other design features, or attacked through various vectors to trick the model into learning or acting incorrectly, or revealing data. In this last class of attacks, the ML model can be interrogated to leak information about sensitive training data. This data, if compromised, could provide an adversary with critical insight into capabilities, intentions, vulnerabilities, and operational plans. In this work, we introduce an automated developmental test and evaluation (DT\&E) pipeline that seeks to proactively quantify the risk of data leakage through model inversion. By rapidly analyzing vast amounts of data, at a speed and scale that far beyond human capabilities, our pipeline can assess susceptibility to model inversion attacks and provide potential risk-level information to stakeholders and mission planners.

\subsection{Model Inversion Attacks}

Model inversion attacks (MIAs) aim to reconstruct a ML model's original training data by exploiting correlations it has between inputs, internal representations, and outputs, such as predictions and confidence scores. A successful inversion attack will reveal private aspects of the training dataset, posing a direct risk to the confidentiality of training data. In addition, an adversary may be able to create a well-performing surrogate model with the reconstructed data, posing additional risks to the model. The specific MIA methodology depends heavily on the adversary’s knowledge and access level, but is typically achieved by optimization. 

The strongest adversary, referred to as a “white-box” attack, will have access to gradients, as in Fredrikson’s original MI-Face algorithm \cite{Fredrikson2015}. The weakest adversary, referred to as a “black-box” attack, will have only access to the model’s outputs (soft or hard labels). Modern MIAs leverage additional tools beyond the optimization technique employed. These include the use of generative techniques such as Generative Adversarial Networks (GANs) that can reconstruct high-fidelity representations of training data in ``black-box'' scenarios that can be superior to the equivalent “white-box” results. Additionally, an attacker can leverage vast amounts of public data to assist in a better attack.

\subsection{Related Privacy Risk Frameworks}

Rigorously quantifying the privacy risk to training data from MIAs is an open area of research. A survey finds only a handful of frameworks that incorporate this attack type---all were designed for assessing the risk from either the membership inference attack, another privacy attack type, or the evasion attack that results in misclassifications from small changes in input data. No frameworks or toolkits were found that develop risk measures for MIAs. No techniques were identified that utilize emerging vision language models (VLMs) to interpret reconstructed data from MIAs. 

\subsection{Identifying Gaps: Our Approach}

The contribution offered here is an automated, modular framework that streamlines the evaluation and replication of MIAs across diverse architectures and datasets. Understanding and quantifying image inversion quality can be inherently challenging and subjective, particularly given the enormous scale of models, images, and classes. A workload of this size is virtually impossible for human analysts. Our DT\&E pipeline tackles this by efficiently processing large amounts of data and ML models to quantify operational risk. Rather than replacing the evaluation model with an MLLM to determine MIA accuracy, we integrate VLMs to enhance the interpretability and analysis of our reconstructed images. We analyze VLM output for a large number of class reconstructions to characterize the purpose of the model’s task for each class. We then use this information and the reconstructed images to train a proxy model that represents an adversary’s ability to perform model stealing; that is, to create models with a similar task to the target but without access to the private data or model architecture. Together, these provide four dimensions to quantify and assess the risk to the private data from MIAs. Two are variants of existing evaluation techniques, and the other two are innovations. We focus on coarse-grained imagery classification tasks where the categories represent a broad range of objects---since current research indicates that VLMs are less capable in fine-grained recognition \cite{chandhok2024response}. While we do not assume knowledge of class labels, reasonable inferences about model output can be made based upon input data formats. With this knowledge, additional insights can be gleaned using public datasets and GANs. When we employ an evaluation model to assess MIA accuracy, we ensure that the training dataset for the evaluation model and the training dataset that was used to train the target model are independent. 

\section{METHODOLOGY}

Our approach focuses on applying MIAs against image classification ML models due to their extensive use in military settings. We design a DT\&E framework that quantifies the inversion risk to a target model through an average risk value (see Fig.~\ref{fig:pipeline}). Assuming different levels of access to the target model in the MIA, we repeatedly apply an appropriate MIA technique to generate multiple reconstructed images for each class. The components of our modular framework are described in detail in Section~\ref{pipeline stages}. The evaluation of our pipeline is described in Section~\ref{eval}. Each of these images is assessed across multiple risk dimensions and the results are aggregated at the class level. The classification accuracy in each risk dimension is calculated from the ground truth and is directly related to the risk score. These risk scores enable comparison of target models to assess which are more vulnerable to the MIA on a per-class resolution. The modular framework and all experiments were completed using PyTorch.

\subsection{Pipeline Stages} \label{pipeline stages}

Our framework integrates VLMs to enhance the interpretability and analysis of our reconstructed images. VLMs have a rich latent space that can enhance inversions with textual descriptions and contextual understanding. This not only removes subjective interpretations, but also offers a more comprehensive evaluation of the privacy risks associated with the reconstructed images. This approach can help identify sensitive information that may be exposed through MIAs and supports the development of better defense mechanisms. The process is shown in Fig.~\ref{fig:pipeline} and is broken down into four stages.

\begin{figure}[t]
    \centering
    \vspace{6pt}
    \includegraphics[width=\linewidth]{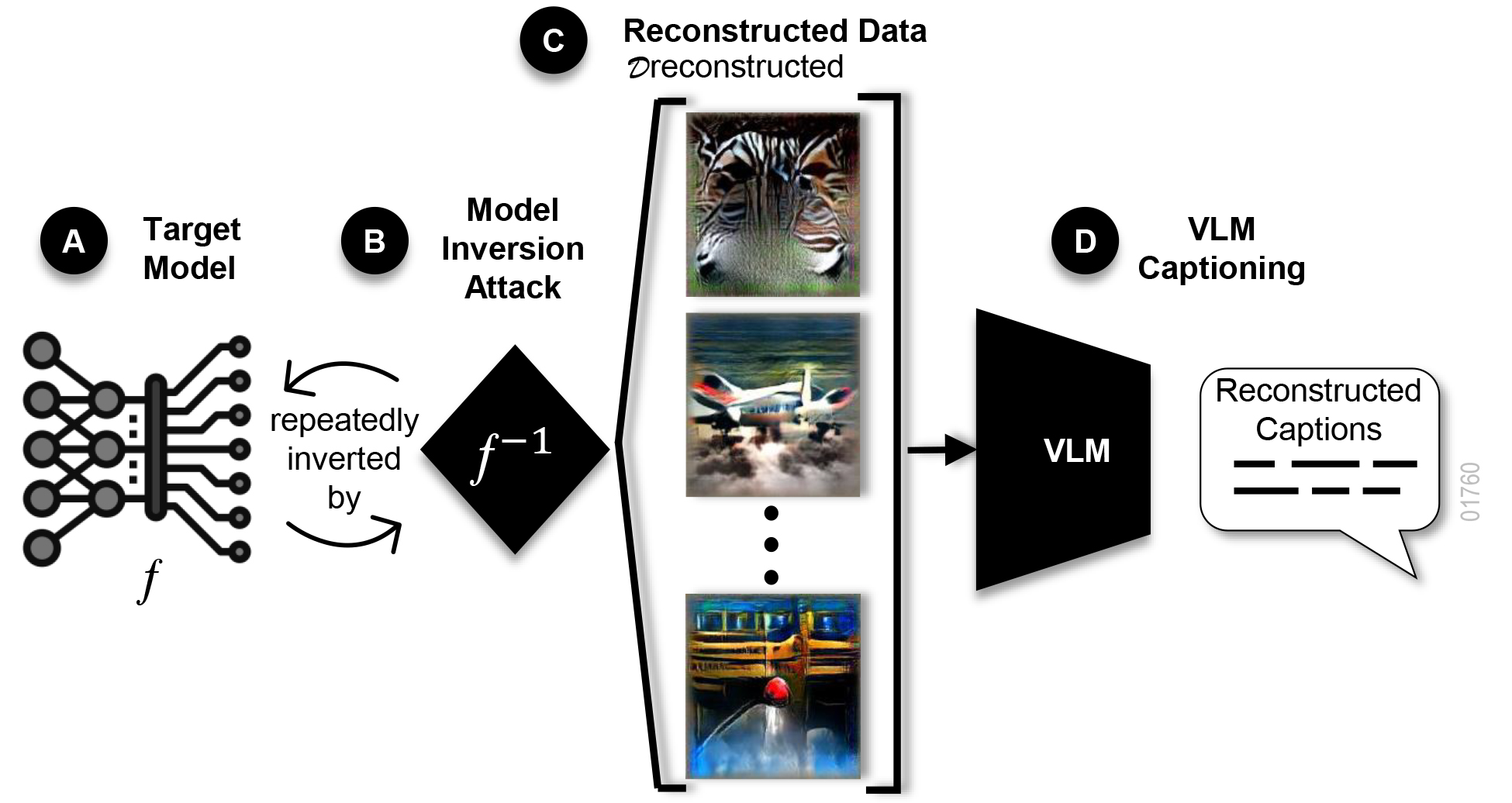}
    \caption{An Automated Model Inversion DT\&E Framework.}
    \label{fig:pipeline}
\end{figure}

\noindent \textbf{A. Target Model:} This is the target model for assessment that will processed. Models are classified as high, medium, or low risk based on its susceptibility to data leakage.\\
\textbf{B. Model Inversion Attack:} MIAs will be employed repeatedly against the target model to produce the required number of image reconstructions for the subsequent risk analysis. In some attacks, a pretrained GAN is used as a prior to generate realistic-looking images from a latent space.\\
\textbf{C. Reconstructed Data:} The pipeline repeatedly creates reconstructed images using the target model and an existing MIA. The output is a reconstructed image that strongly activates a specific internal layer of the target model. The set of all such images generated through repeated inversions in an experiment is denoted $D_{reconstructed}$. These may appear visually realistic due to the effectiveness of the MIAs. Although the GAN contributes a generative prior, visual interpretability is not essential to the pipeline’s function.\\
\textbf{D. VLM Captioning:} A textual description of every reconstructed image is generated with a VLM, offering additional insights into model behavior, the training data characteristics, and features that influence model decisions. Prompt engineering may be applied when VLM-supported.\\

\subsection{Evaluation Metrics} \label{eval}
For each target model, we assess the reconstructed images using four evaluation metrics: quality loss, feature loss, label loss and model stealing loss. Fig.~\ref{fig:evaluation_pipeline} provides the steps taken to determine and use these evaluation metrics.

\noindent \textbf{1) Quality Loss:} Using the inferred class labels, an adversary may generate an evaluation model using public data. The pipeline incorporates an “Evaluation Model” \cite{park2019}, that represents an adversary's ability to create models with similar tasks to the target, but does not assume data access or model architecture knowledge. Thus, the model is trained using data that is disjoint from the target model's training dataset. In DT\&E, the target model’s test dataset can be used to obtain a model with similar performance. Through this model, the reconstructed images are inferred and assessed for classification accuracy. Higher accuracy indicates a higher information loss of the training data through the target model. We term this information loss as “quality loss” since an adversary can understand how well the target model performs its classification task with this metric.\\
\textbf{2) Feature Loss:} ``Feature loss” measures an adversary’s ability to infer possible class labels for each model output node. A VLM generates a textual description for each reconstructed image, providing additional insights into model behavior, training data, and the features that influence model decisions. In our framework, we apply zero-shot text classification \cite{facebook/bart-large-mnli} to the image descriptions, generating a list of potential class labels. These are compared to the known classes to compute classification accuracy. Higher accuracy indicates greater information leakage from the training data through the target model.\\
\textbf{3) Label Loss:} “Label loss” refers to the emergence of high-confidence class labels that reveal the underlying purpose of the model’s task. For each model output class, zero-shot classification is performed using a VLM, the corresponding group of reconstructed images, and the list of potential labels generated during the “feature loss” step. For benchmarking, the class labels used are from known classes, and we calculate the accuracy of obtaining the correct class label. Higher accuracy indicates a higher information loss of the training data through the target model. For realistic attack scenarios where the classes may be unknown, we evaluate class accuracy with incorrect class labels.\\
\textbf{4) Model Stealing Loss:} ``Model stealing loss" captures how effectively an adversary can combine reconstructed data and public data to create their own model with similar performance for the same classification task. The evaluation model is fine-tuned using the reconstructed images to create a proxy model, ideally with similar task performance. To measure this ability, the target model’s training dataset is inferred through the proxy model and assessed for classification accuracy. Higher accuracy indicates greater information leakage of the training data through the target model, which indicates a greater ability for an adversary to steal the target model. This metric, also known as the Knowledge Extraction Score \cite{he2024becareful}, measures the extracted discriminative information about distinct classes.

\begin{figure}[t]
    \centering
    \vspace{6pt}
    \includegraphics[width=\linewidth]{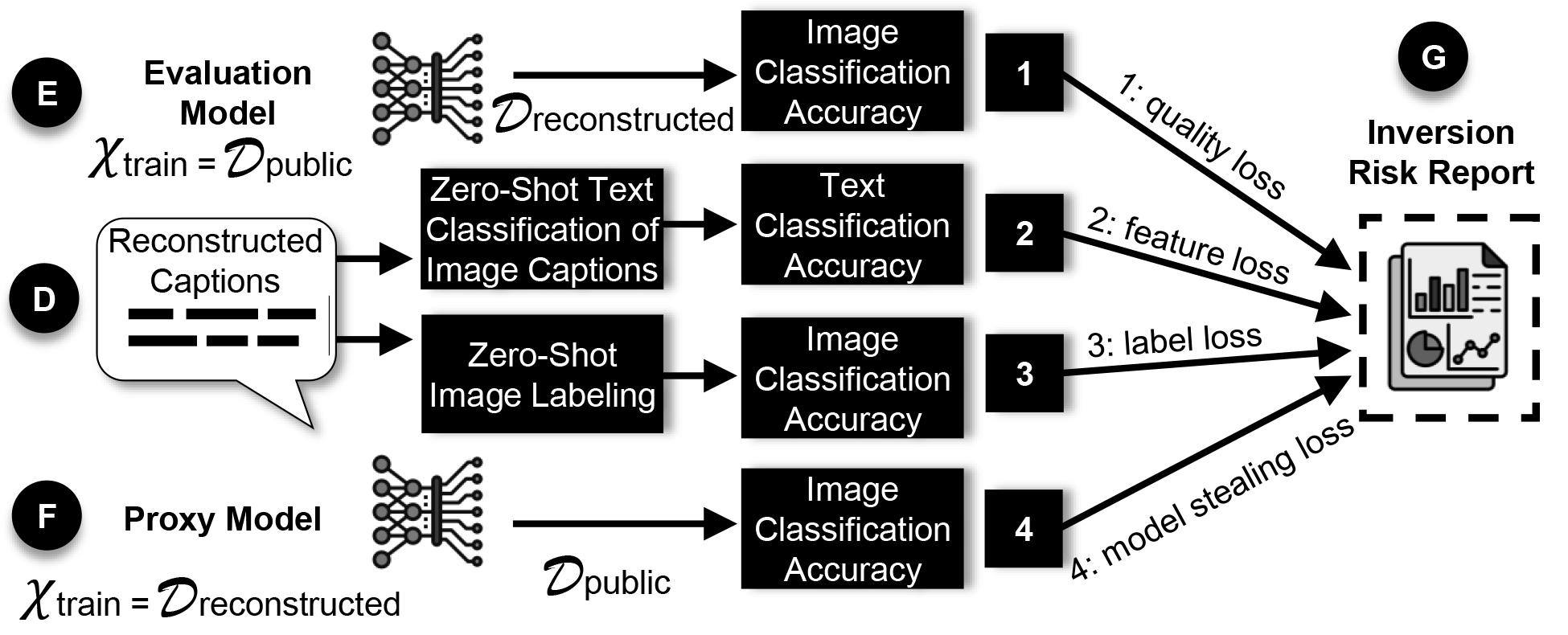}
    \caption{Pipeline Evaluation Metrics.}
    \label{fig:evaluation_pipeline}
\end{figure}

\noindent \textbf{D. VLM Captioning:} Reconstructed image captions are used in two ways for pipeline evaluations: zero-shot text classification and zero-shot image classification. Zero-shot text classification is performed on the VLM output text descriptions, providing a list of potential class label names the model is trained against. Zero-shot image classification is performed on each reconstructed image, providing a label for each image. Together, these give information about how the VLM interprets the reconstructed image.\\
\textbf{E. Evaluation Model:} Using the inferred class labels, an evaluation model is generated using public data, $D_{public}$. This represents an adversary's ability to create models with a task similar to the target model's. Through this model, the reconstructed images are inferred and assessed for classification accuracy. We employ a different architecture for the evaluation model than for the target model. For evaluation model training, we use a dataset that is disjoint from the target model's training dataset.\\
\textbf{F. Proxy Model:} The evaluation model is fine-tuned using the reconstructed images to create a proxy model, ideally with similar task performance. To measure this ability, the target model’s training dataset is inferred through the proxy model and assessed for classification accuracy.\\
\textbf{G. Inversion Risk Report:} In the final step of the pipeline, the four risk metrics below are calculated and averaged (with equal weight herein) to produce a single privacy risk score, weighted composite accuracy loss (WCAL), where $k$ is the respective risk dimension, $\omega_k$ is the risk weight given to the $k^{th}$ dimension, and $r_k$ is the loss value, as expressed by: 
\begin{equation}
\text{WCAL} = {\sum_{k=1}^{4} (\omega_k \times r_k)}, \text{   subject to   }  {{\sum_{k=1}^{4} \omega_{k} = 1}}
\end{equation}

This WCAL score provides an overall risk value for the MIA against the target ML model. By adjusting the weights, specific risk dimensions can be emphasized or de-emphasized. This score is then used to bin a model into low ($<55\%$), medium ($55\%-65\%$), and high ($>65\%$) risk categories.

\subsection{Experimental Design}

We evaluate our framework using a suite of configurations to test its effectiveness across different image datasets, target model architectures, MIA techniques, numbers of reconstructed images, and VLM versions. We first conduct experiments against a single dataset to analyze pipeline performance under varying configurations. In all experiments, InceptionV3 \cite{inceptionv3} is used for the evaluation of reconstructed image quality via the quality loss metric. We then apply the best and worst weighted composite loss configurations to a new dataset and observe pipeline performance. In Experiment 1, we assess how the number of reconstructed images and the target model architecture affect composite risk. We evaluate results using 10, 50, and 100 image reconstructions. For the target models, we utilized three widely adopted architectures: VGG16 \cite{vgg16}, MobileNetV2 \cite{mobilenetv2}, and ResNet50 \cite{resnet50}. Each of these models was trained from scratch and achieved competitive performance. 

In Experiment 2, we assess the impact of different MIA types upon composite risk. To reconstruct images from the target models, we implemented two MIAs: Feature Visualization (FV) with a GAN prior \cite{zeiler2014visualizing} and Generative Model Inversion (GMI) \cite{zhang2020secret}. FV creates a natural image that maximally activates the target class node via constrained optimization with a GAN image prior. The GMI approach trains a GAN to generate samples resembling the target model’s training data, using the target model as the discriminator. In essence, FV uses a GAN to help understand a model's learned concepts in a general way, while a GMI attack attempts to exploit the model to recover specific, private-training instances, often employing generative techniques to do so. 

In Experiment 3, we evaluate the impact of different VLMs on zero-shot image labeling and text classification. Specifically, we use CLIP \cite{clip}, BLIP2 \cite{blip2}, and InstructBLIP \cite{instructblip}. In Experiment 4, we assess the effect of using different datasets on model inversion performance. Experiments 1 through 3 were conducted using the STL-10 image recognition dataset \cite{coates2011analysis}. STL-10 contains ten distinct image classes that were acquired from labeled examples on ImageNet \cite{deng2009imagenet}, as shown in Fig.~\ref{fig:evaluation_pipeline}. 
In Experiment 4, we use a vehicle classification dataset \cite{labv2_military_vehicles_2023} designed to train and evaluate ML models that distinguish between military vehicle categories, such as tanks, air defense, and self-propelled artillery.

\textbf{Impact of Public Knowledge:} Distilling prior knowledge and properly incorporating it into the MIA algorithm is important to attack success. Our method incorporates prior knowledge from public datasets into the pipeline via the GAN. GAN training is a separate step and involves an additional wide range of confirmation parameters beyond those of the pipeline. Although our experiments did not focus on finding the optimal parameters for GAN training, we list results from several different configurations and propose this as an avenue for future research.

\section{RESULTS AND ANALYSIS}
All experiments were conducted within a containerized environment on a dual-GPU system equipped with two NVIDIA Tesla P100-PCIE-16GB GPUs. The system was configured with NVIDIA Driver 555.42.02 and CUDA 12.5. For consistency and reproducibility, we used Docker with the official NVIDIA PyTorch container (rel-23-12), which includes PyTorch 2.2.0. 
The MIA Toolbox was used for the GMI experiments and Lucent \cite{lucent} for the FV experiments. Each experiment was run 30 times per configuration, with results averaged across trials and a 95\% confidence interval computed.

\subsection{Experiment 1: The Impact of the Number of Model Inversions and Target Model Architecture}


This experiment assesses model vulnerability using varying numbers of reconstructed images of 10, 50, and 100 per class. We also vary the target model architectures to observe their impact on risk, using VGG16, MobileNetV2, and ResNet50. Feature Visualization with an ImageNet GAN as the image prior serves as the inversion technique. The results for feature loss, label loss, quality loss, model stealing loss, and weighted composite accuracy loss are recorded in 
Fig.~\ref{fig:loss_graphs}, which reveals several key observations. First, the label losses are significantly higher than the associated quality and model stealing losses. This is also true for the VGG16 feature loss. This confirms that our privacy loss metrics are measuring information loss not previously observed. Second, increasing the number of reconstructed images per class does not significantly affect any of the loss values. This suggests that using fewer reconstructions per class can offer similar result quality with improved computational efficiency. Third, the quality loss trend indicates that VGG reconstructions capture class characteristics the least accurately, while MobileNetV2 reconstructions are the most accurate. Lastly, model stealing loss tends to decrease with more reconstructions for all models, implying that as more image reconstructions are used in the training of the proxy model, the proxy model becomes overfit with no additional gains from further inversions.

\begin{figure}[t!]
    \centering
    \vspace{6pt}
    \includegraphics[width=\linewidth]{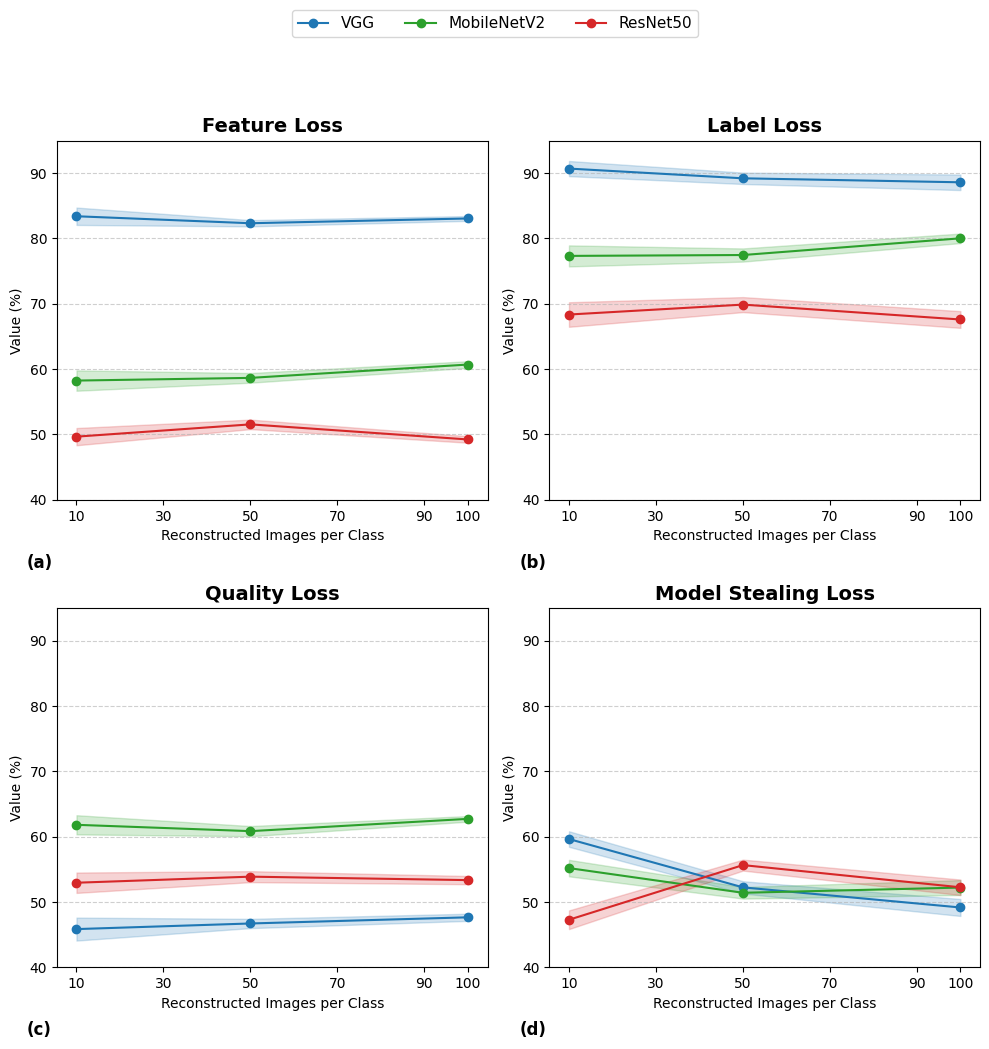}
    \caption{Metric Trends Across Image Counts and Architectures.}
    \label{fig:loss_graphs}
\end{figure}

\begin{figure}[t!]
    \centering
    \includegraphics[width=0.9\columnwidth]{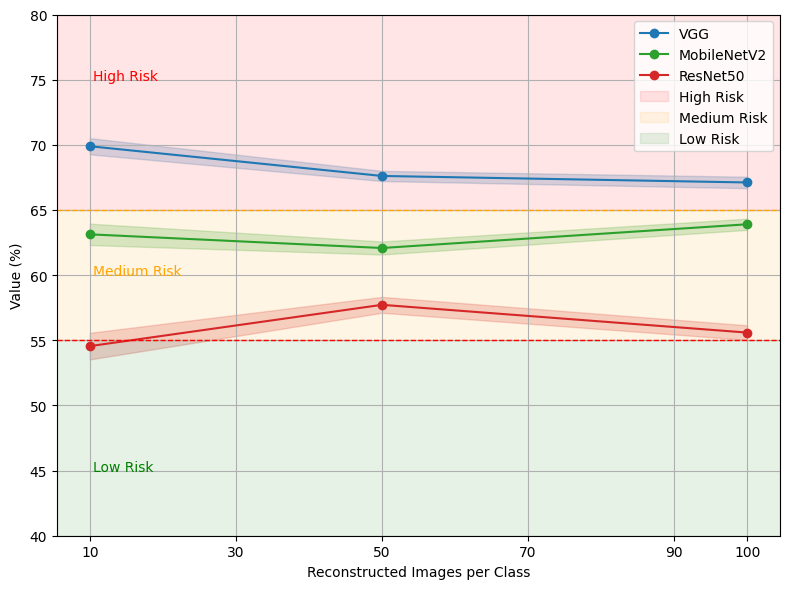}
    \caption{Weighted Composite Accuracy Loss Across Image Counts.}
    \label{fig:weighted_composite_loss_graph}
\end{figure}

Fig.~\ref{fig:weighted_composite_loss_graph} illustrates the WCAL for each model architecture as the number of reconstructions per class increases. The shaded regions denote relative risk levels: high risk (\(\geq65\%\)), medium risk (55–65\%), and low risk (\(\leq55\%\)) and provide a qualitative assessment of overall model exposure. VGG consistently exhibits the highest composite loss, designating it as high-risk. MobileNetV2 is designated as medium-risk and ResNet50 as low-risk. These results suggest that VGG16 is consistently the most vulnerable to MIA data leakage, while ResNet50 exhibits stronger resistance. This implies that model architecture does have an impact on the amount of data leakage and should be considered before model deployment. This data also supports the conclusion that increasing the number of reconstructed images does not increase risk.

\subsection{Experiment 2: The Impact of Different MIAs}

This experiment investigates the impact upon weighted composite risk of two different MIAs: FV with a GAN prior and GMI. The GMI approach involves training a GAN to generate samples that look like the training data of the target model, using the target model as a discriminator. Based upon the results of Experiment 1, the target model was chosen to be a VGG16 model. The model's train accuracy was 98.4\% and test accuracy was 95.2\%. Ten reconstructed images per class were produced with the GMI attack. While optimizing the GAN configuration was not a focus of our experiments, we prioritized GANs that produced reconstructed images with visually salient features---even if they were not fully human-interpretable. To support this goal, we used three configurations, referred to as GMI configurations A, B, and C, each resulting in visually salient reconstructions. The three GMI attack configurations show a clear progression in computational intensity and duration. Configuration A was the fastest, running for 5,000 iterations with a combined score function threshold of 0.7, completing the attack in 43,760 seconds. Configuration B increased the workload to 7,500 iterations and raised the selectivity threshold to 0.8, which more than tripled the attack time to 133,304 seconds. Configuration C was the most resource-intensive, extending to 10,000 iterations while maintaining the 0.8 threshold, resulting in the longest attack time of 173,021 seconds. This demonstrates a direct correlation where a higher number of iterations, and to a lesser extent a stricter selection threshold, substantially increase the overall time required to execute the attack. Fig.~\ref{fig:stl10_images_reconstructions} depicts representative real STL-10 images and reconstructed images from FV and GMI Configuration C for each class.


\begin{figure*}[t]
    \vspace{6pt}
    \includegraphics[width=\linewidth]{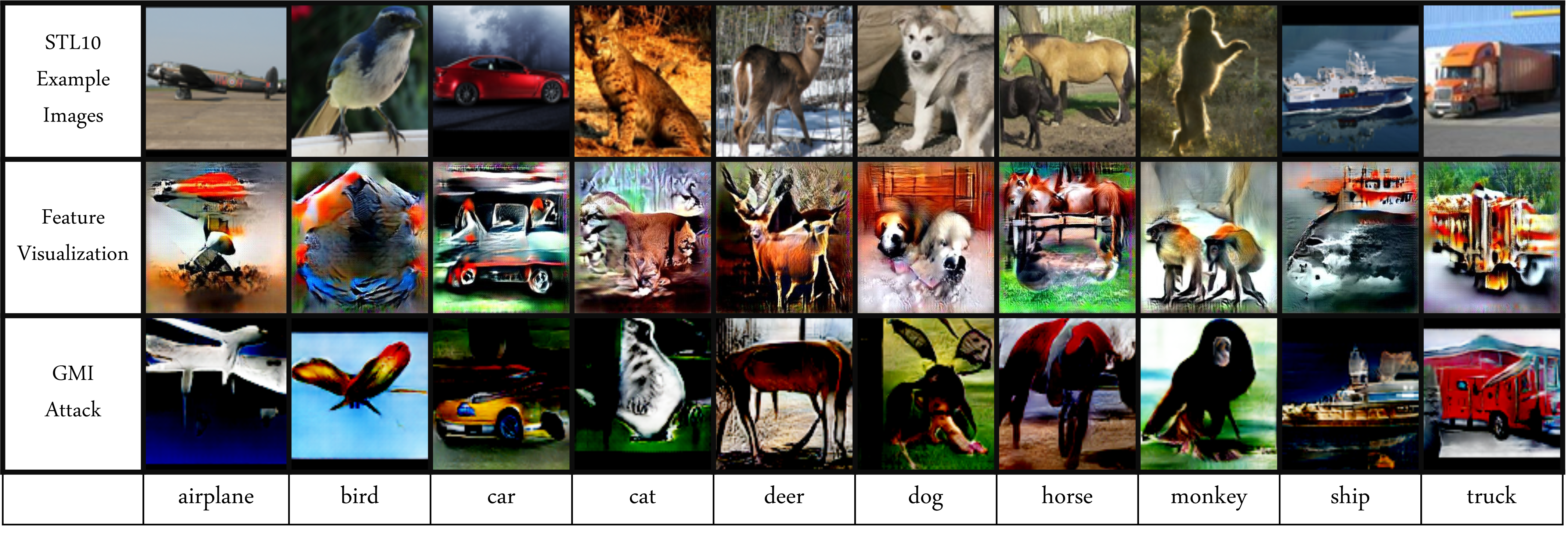}
    \caption{Representative Data Reconstructions for STL-10 Classes.}
    \label{fig:stl10_images_reconstructions}
\end{figure*}


The results illustrated in Fig.~\ref{fig:experiment2_results_graph} show that configuration C 
obtained the highest loss values across all metrics and obtained a weighted composite accuracy loss of 60.72\%. When compared to the FV results 
in Fig.~\ref{fig:loss_graphs}, configuration C's GMI results have higher loss values for quality and model stealing loss. The weighted composite accuracy loss for FV is higher than those of GMI, due to the higher label loss in FV that pushed the average higher. This implies that with additional time and resources to implement GMI and optimize configuration, a higher privacy loss may be obtained. 

\begin{figure}[t!]
    \centering
    \includegraphics[width=.9\linewidth]{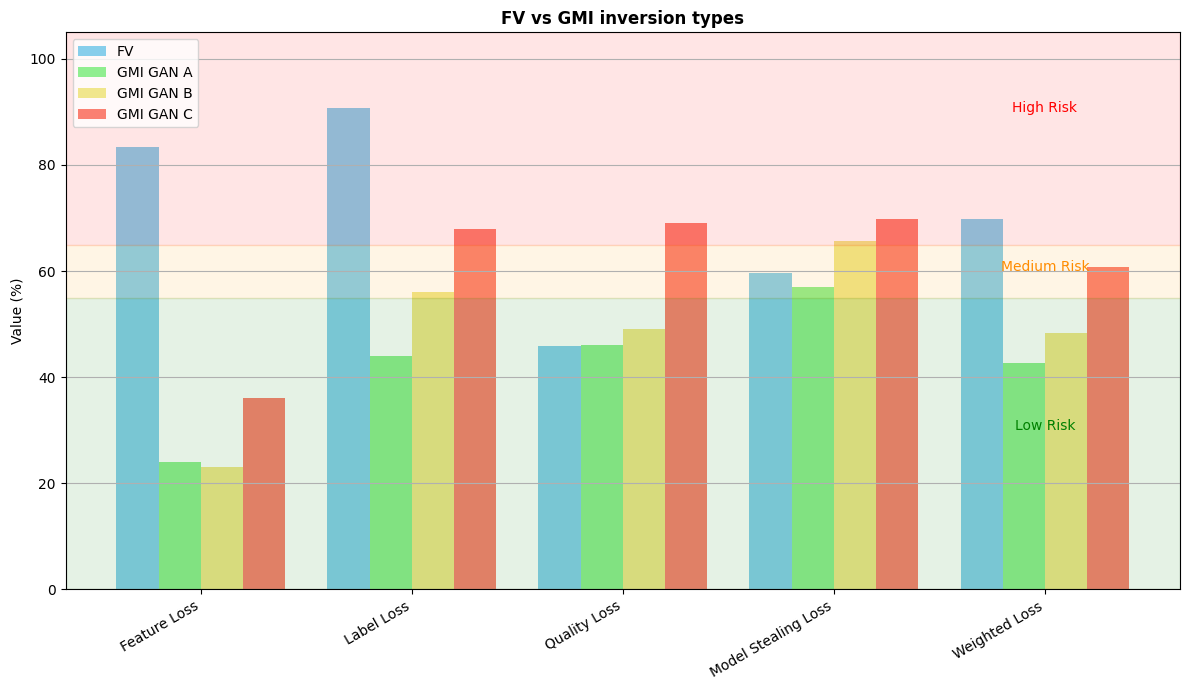}
    \caption{Different Model Inversion Attack Results.}
    \label{fig:experiment2_results_graph}
\end{figure}

\subsection{Experiment 3: The Impact of VLM Type}

Experiment 3 focuses on evaluating the effectiveness of two different VLMs, BLIP2 and InstructBLIP, in describing the reconstructed images. BLIP2 (Bootstrapping Language-Image Pretraining) is a VLM designed for image-to-text generation. It uses a two-stage design. A frozen vision encoder first extracts visual features which are routed through a lightweight Q-Former that transforms those features into an output the language model can interpret. The language model then produces relevant text based on the identified features. This design allows BLIP2 to excel in generating concise captions based on visual content without additional prompts. InstructBLIP builds on BLIP2 by incorporating instruction tuning, making it more responsive to prompting and reasoning. This enables InstructBLIP to produce captions that go beyond surface-level descriptions and demonstrate contextual knowledge, unlike BLIP2 which only generates general captions. 

In this experiment, all three target model variants (VGG, MobileNetV2, and ResNet50) are employed and all trials use the FV MIA algorithm. $D_{reconstructed}$ consists of 10 reconstructed images per class which are inferred through both BLIP2 and InstructBLIP to produce captions. These captions are expected to differ, possibly resulting in improved evaluation metrics. 
Fig~\ref{fig:vlm_comparison_graph} presents the resulting feature loss, label loss, and weighted composite accuracy loss for each VLM–model pairing, highlighting the VLM caption quality.


In the DT\&E pipeline, the VLM does not impact either quality loss or model stealing loss, so we do not report these metrics in this experiment. 
Both VLMs produced similar results across all architectures. BLIP2 slightly outperformed across each risk metric with VGG, and InstructBLIP showed better performance with MobileNetV2 and ResNet.

\begin{figure}[H]
    \centering
    \includegraphics[width=\linewidth]{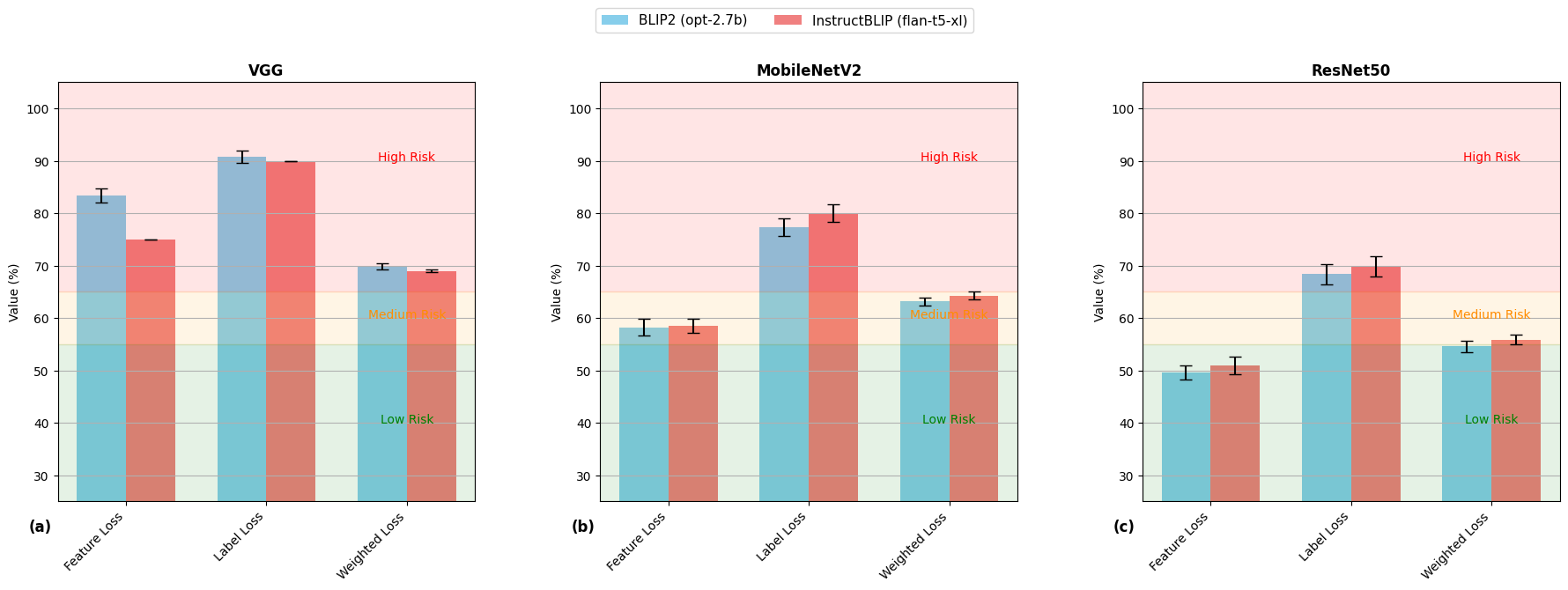}
    \caption{VLM Comparison Across Model Architectures.}
    \label{fig:vlm_comparison_graph}
\end{figure}

With VGG, InstructBLIP yielded a feature loss of 75.0\% and a label loss of 90.0\%. The WCAL reached 69.0\%, placing this configuration near the lower bound of high risk. On MobileNetV2, InstructBLIP achieved a feature loss of 58.52\% and a label loss of 80.05\%. The WCAL was 64.27\% ±0.73, hovering just below the high risk threshold. For ResNet50, InstructBLIP produced a feature loss of 50.98\% and a label loss of 69.84\%. The weighted average accuracy settled at 55.88\%, right at the boundary between low and medium risk. InstructBLIP could be further guided to exploit a model if prior knowledge of a model or dataset is known. For example, if an attacker knows the objects that are likely present in the training data, targeted prompts can be used to guide InstructBLIP's caption generation leading to a higher potential for data leakage and model stealing.

\subsection{Experiment 4: Application to a Military Dataset}
In this experiment, we investigate our model inversion assessment pipeline using a different dataset, one with a direct military application. We use a military vehicle classification dataset \cite{kricheli2024error}, which is imbalanced with 24 fine-grained and 7 coarse-grained classes of military vehicles. The model task is to classify each image into one of the military class labels. It has been documented that VLMs show a drop in performance in fine-grained categories due to vision language projection and alignment with the language decoder, but decoders within VLMs that are instruction fine-tuned can bridge this performance gap \cite{chandhok2024response}. We use the coarse-grain resolution hierarchy, as we do not anticipate the stock BLIP2 VLM to successfully differentiate between military vehicle versions, i.e., T-62, T-72, T-90 tanks. To further facilitate these trials, we only utilize 5 of the 7 classes in this experiment: Air Defense, BMP, BTR, Self-Propelled Artillery, and Tank. The BMD and MT-LB categories were unchanged from fine-grain to coarse-grain classification, so they were omitted from this experiment. The first trial for this experiment uses the general parameters used with the previous STL-10 experiments, and the second trial utilizes a pipeline fine-tuned for the military vehicle dataset. 
The target model is trained with a VGG16 architecture and a NearMiss undersampling technique (version = 1, neighbors = 3 and feature size = 32 x 32) is used in the dataset, since it is an imbalanced dataset \cite{zhang2003knn}. We use FV with an ImageNet GAN for image priors as the inversion technique to generate 50 reconstructed images per class. Based on the research carried out with the dataset, we use ViT-B/16 \cite{vit-b-16} for the evaluation and proxy models architecture instead of InceptionV3. For the first trial, we use a subset of the dataset containing five classes, with VGG16 as the target model architecture. This model achieved a training accuracy of 79.39\% and a test accuracy of 69.77\%. The evaluation model demonstrated moderate performance with a training accuracy of 80.18\% and a test accuracy of 69.77\%. The proxy model used in this experiment showed a training accuracy of 53.12\% but a significantly lower test accuracy of 24.85\%. The results of these trials are recorded in the first row of Table~\ref{fig:military_vehicles_results}. In general, lower loss values are observed compared to previous STL-10 experiments. We attribute this to the lower target model training and test accuracies in these trials. We also observe that the VLM-based metrics, label loss and quality loss, are lower than the STL-10 experiments because, even though we chose the coarse-grain classification configuration for the military vehicle dataset, the data are still more fine-grained in nature compared to STL-10. The model stealing loss is lower because the accuracy of the proxy model is lower compared to that of the target model. The quality loss values are high because the process of creating the reconstructed images saves images that are correctly categorized against the target model. While this results in good accuracy against the evaluation model, the difference in loss scores reinforces the observation that this MIA evaluation does not capture true visual semantics \cite{ho2025revisiting}.

\begin{table}[t!]
     \centering
     \caption{Experiment 4 Results.}
     \includegraphics[width=.9\linewidth]{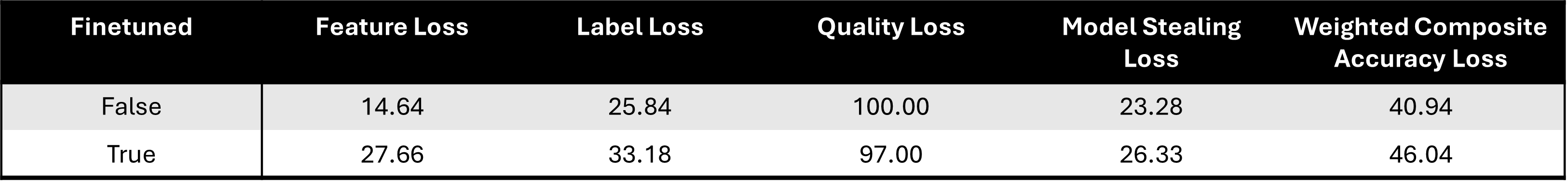}
     \label{fig:military_vehicles_results}
 \end{table}

To demonstrate how distilling and incorporating prior knowledge can improve MIA success, we fine-tuned the GAN used as the prior in the FV attack, as well as the BLIP and CLIP models used for feature loss and label loss evaluation. Prior knowledge utilizes public datasets in practical MIA, but to capture the highest-risk assessment of the models and dataset, we utilize training data for the following experiments. To fine-tune the GAN, we implemented a DCGAN trained on the military vehicle dataset. The goal was to generate realistic images of vehicles such as tanks, BTRs, and air defense systems to serve as priors for FV. The GAN was trained with the same data subset and NearMiss undersampling as the target model. 
To fine-tune the VLMs, the military vehicles dataset needed to be converted into a local format suitable for image captioning tasks. The dataset was filtered to include the five classes, as noted above. We generate a simple, template-based caption for each image, such as ``a photo of a tank,'' and record this text. The final output is a self-contained local dataset, structured for easy loading with the Hugging Face datasets library, ideal for fine-tuning vision language models. We fine-tuned two pretrained vision language models---BLIP for feature loss evaluation and CLIP for label loss evaluation---using the military vehicles dataset. Both models were fine-tuned using Parameter-Efficient Fine-Tuning (PEFT) LoRA (Low-Rank Adaptation) method to reduce memory consumption and training time by freezing most of the original model weights and updating only a small subset of parameters. 
The datasets were prepared by processing the images and generating class-specific text 
which were tokenized and formatted with the respective model's processor (BLIPProcessor or CLIPProcessor). 
For the feature loss evaluation, we fine-tuned BLIP \cite{blip} to generate captions for military vehicle images. The fine-tuned BLIP model was evaluated on test military vehicle images with captioning accuracy of 70.2\% compared to 36.3\% of the base BLIP model. For label loss evaluation, we fine-tune a pretrained CLIP 
model for an image classification task using the military vehicles dataset. Instead of conventional classification, it leverages CLIP's ability to learn joint representations of images and text by creating descriptive text prompts for each class 
and teaching the model to associate images with the corresponding text descriptions. 
The fine-tuned CLIP model was evaluated on the test military vehicles images with a 36\% classification accuracy compared to an 11\% classification accuracy of the base CLIP model. For the second run using a subset of the military vehicle dataset, the VGG16 model architecture was used. This model achieved a training accuracy of 79.1\% and a test accuracy of 69.32\%. The same evaluation model was used as the first run which achieved a training accuracy of 80.18\% and a test accuracy of 69.77\%. The proxy model used in this experiment showed a training accuracy of 62.82\% but a test accuracy of 28.73\%. Both of these values increased with the use of the military vehicles GAN as a prior for FV. The results of these trials are recorded in the second row of Table~\ref{fig:military_vehicles_results}. While these results still have lower loss values when compared to previous STL-10 experiments, they exhibit an increased loss from the first run without fine-tuning. These increases show that the fine-tuning of the GANs and VLMs allow more information to be leaked by the target model. For these experiments, the label loss value was calculated using the known model classes as the candidate labels.

\section{CONCLUSION}

This work presented an automated, scalable DT\&E pipeline for assessing the risks posed by MIAs to ML models, along with new metrics to measure this privacy loss. By integrating VLMs and leveraging modularity, the pipeline provides a comprehensive framework to assess the susceptibility of ML models to data leakage. The results demonstrate that the pipeline effectively quantifies privacy risks across multiple dimensions, including quality loss, feature loss, label loss, and model stealing loss, offering a weighted composite risk score to guide decision-making. The key findings highlight the importance of these additional risk dimensions, as current evaluation metrics do not capture visual context information. 

Our findings indicate that our framework enables model architecture, MIA techniques, and dataset characteristics to determine the extent of privacy loss. For example, VGG16 models consistently exhibited a higher vulnerability to MIAs compared to ResNet50, while the use of GMI attacks revealed higher privacy risks than FV attacks under certain configurations. Additionally, the integration of fine-tuned VLMs and GANs tailored to specific datasets, such as military vehicle classification, further enhanced the pipeline's ability to assess and expose privacy risks. 
The modularity of the pipeline ensures adaptability to diverse datasets and ML tasks, making it a valuable tool for both research and operational applications. By aligning evaluation and proxy models with the target model's tasks and incorporating fine-tuned VLMs, the pipeline can be extended to new domains, including those with high-stakes applications like military technology. In addition, the ability to automate and scale the assessment process addresses the challenges posed by the large number of ML models and datasets that require evaluation. 

Future research could incorporate other risk dimensions such as those from new VLM types, i.e. Google Gemini, to score inversions. Also, additional research into model forensics and dynamic analysis may add new risk dimensions. In conclusion, this work advances the state-of-the-art in model inversion risk assessment by providing a scalable, automated, and modular framework. It offers a clearer understanding of vulnerabilities in ML models, enabling stakeholders to make informed decisions to mitigate privacy risks and improve the security of ML-enabled AI systems.

\section*{Acknowledgments}
This research was sponsored by the United States Military Academy (USMA) under Cooperative Agreement No. W911NF-25-2-0034. The views and conclusions expressed in this paper are those of the authors and do not reflect the official policy or position of the U.S. Military Academy, U.S. Army, U.S. Department of Defense, or U.S. Government.

\bibliographystyle{IEEEtran}
\bibliography{model_stealing} 

\end{document}